\documentclass[12pt,english,floatfix,showkeys,superscriptaddress,aps,prd,preprint]{revtex4}

\usepackage[utf8]{inputenc}
\usepackage[english]{babel}
\usepackage[T1]{fontenc}
\usepackage{lmodern}
\usepackage{amsmath}
\usepackage{amssymb}
\usepackage{graphicx}
\usepackage{float}
\usepackage{esint}
\usepackage{dcolumn}
\usepackage{csquotes}
\usepackage{color}
\usepackage{slashed}
\usepackage{simplewick}
\usepackage{subcaption}

\usepackage{hyperref}
\hypersetup{
    colorlinks,
    citecolor=blue,
    filecolor=green,
    linkcolor=purple,
    urlcolor=red,
    breaklinks=true,
    urlcolor=[rgb]{0.0,0.0,1.0},
    linkcolor=[rgb]{0,0.5,0.9}
}

\begin{document}

\title{Fermionic Casimir Energy in Horava-Lifshitz Scenario}

\author{E. R. Bezerra de Mello}
\email{emello@fisica.ufpb.br}
\affiliation{Departamento de Física, Universidade Federal da Paraíba, Caixa Postal 5008, 58051-970, João Pessoa, Paraíba, Brazil.}

\author{M. B. Cruz}
\email{messiasdebritocruz@servidor.uepb.edu.br}
\affiliation{Centro de Ci\^encias Exatas e Sociais Aplicadas, Universidade Estadual da Para\'iba, \\ R. Alfredo Lustosa Cabral, s/n, Salgadinho, Patos - PB, 58706-550 - Brazil.}

\date{\today}

\begin{abstract}
In this work, we investigate the violation of Lorentz symmetry through the Casimir effect, one of the most intriguing phenomena in modern physics. The Casimir effect, which represents a macroscopic quantum force between two neutral conducting surfaces, is widely regarded as a triumph of Quantum Field Theory. In this study, we present new results for the Casimir effect, focusing on the contribution of mass associated with fermionic quantum fields confined between two large parallel plates, in the context of Lorentz symmetry violation within the Horava-Lifshitz formalism. To calculate the Casimir energy and pressure, we impose a MIT bag boundary condition on the two plates, compatible with the higher-order derivative term in the modified Dirac equation. Our results reveal a strong influence of Lorentz violation on the Casimir effect. We observe that the Casimir energy is significantly affected, both in intensity and sign, potentially resulting in a repulsive or attractive force between the plates, depending on the critical exponent associated with the Horava-Lifshitz formalism.
\end{abstract}

\maketitle
\section{Introduction} \label{int}

The Casimir effect refers to the interaction between neutral objects due to the boundary conditions imposed on the quantum fluctuations of fields at zero point. In the context of Quantum Field Theory (QFT), the vacuum state is understood as the sum of all states of quantum oscillators in their ground states. Therefore, the energy associated with the quantum vacuum results in an infinite value. In QFT, various phenomena are associated with the quantum vacuum, with the Casimir effect being one of the principal. This phenomenon is not only significant from a fundamental standpoint but also holds great technical importance in micro and nano mechanical systems.

In 1948, H. B. Casimir theorized the Casimir effect \cite{Casimir:1948dh}, originally considering the vacuum state associated with the electromagnetic field confined between two large, parallel, conducting, and neutral plates. Due to the boundary conditions imposed on the quantum field at the plates, only vacuum excitations with specific wavelengths are allowed between them. In this way, the Casimir energy can be obtained by subtracting the energy associated with these quantum fluctuations from the energy of the vacuum state in the absence of boundaries. This procedure is called renormalization. Using this method, Casimir found an attractive force between the plates, given by:
\begin{equation} \label{cas_forc}
    F = - A \frac{\pi^2 \hbar c}{240 a^4}  \  ,
\end{equation}
where $A=L^2$ represents the area of the plates and $a$ denotes the distance between them, with $a<<L$. This phenomenon has been confirmed experimentally ten years later by  M. J. Sparnaay \cite{Sparnaay:1958wg}; however, due to the low precision, many others experiments were carried out in the 90s, confirming the Casimir effect with a high degree of precision \cite{van1978phgm, lamoreaux1997sk, Mohideen:1998}.

As previously mentioned, the Casimir effect was initially associated with the quantum vacuum for electromagnetic fields. However, in principle, this phenomenon can also occur for other types of relativistic quantum fields, such as scalar and fermionic fields, when subjected to specific boundary conditions. Studies on the Casimir effect considering scalar quantum fields have been conducted in various contexts, including different boundary conditions, temperature, and Lorentz symmetry violation \cite{Khelili:2011pv, Erdas:2013jga, Rego:2013efa, Erdas:2013dha, Cruz_2017,deMello:2022tuv, Vedl:2023izj, Borquez:2023ajx}. In the context of fermionic quantum fields, several studies have been developed for different scenarios \cite{Elizalde:2002wg, Bulgac:2004kk, Queiroz:2004wi, Saharian:2008sz, Teo:2015moa, Teo:2015raa, Grigoryan:2016lcr, Cruz:2018thz, Bellucci:2019ybj,Steinhardt:1979ag,daSilva:2019iwn}. Typically, these studies employ the MIT bag model \cite{Steinhardt:1979ag, Barnhill:1979nx}, originally used to confine quarks in hadrons, assuming a spherical shape for them, as a boundary condition.

The standard QFT represents a successful integration of Special Relativity and Quantum Mechanics, maintaining Lorentz symmetry invariance. However, unifying General Relativity with Quantum Mechanics remains one of the greatest challenges in contemporary physics, with Loop Quantum Gravity \cite{Ashtekar:2007px, Gambini:2011nx} and String Theory \cite{Kostelecky:1988zi, Kostelecky:1995qk} as the leading candidates. Currently, there are two well-known scenarios for introducing Lorentz symmetry breaking. One is the Kostelecky-type model \cite{Kostelecky:2003fs}, where Lorentz-breaking terms are proportional to a small vector (or tensor). The other is the Horava-Lifshitz (HL) model \cite{Horava:2009uw}, where Lorentz symmetry breaking is introduced by considering different orders of derivatives in space and time, resulting in spacetime anisotropy. Although this violation of Lorentz symmetry is thought to occur only at Planck energy scales \cite{Moffat:2009ks}, residual signatures of this spacetime anisotropy may be observable in low-energy phenomena, such as in the study of Casimir effects.

Motivated by the importance of studying a possible violation of Lorentz symmetry, given its relation to fundamental aspects of Quantum Gravity, our aim is to use the Casimir effect as an experimental laboratory. With the advancement of highly precise experimental techniques, we believe in the feasibility of investigating aspects related to the violation of Lorentz symmetry. In this context, the analysis of Casimir energy associated with a massless scalar quantum field confined between two large plates within the HL formalism was explored in \cite{Petrov} and \cite{Ulion:2015kjx}. In these works, Dirichlet, Neumann, and mixed boundary conditions were imposed on the field. For the massive scalar quantum field, this investigation was extended in \cite{Maluf_2020}, representing a significant advancement over previous analyses. The analysis of Casimir effects for massless fermionic quantum fields was conducted in \cite{daSilva:2019iwn}, where the MIT bag boundary condition was imposed on the field between the two large parallel plates. In both cases, for scalar and fermionic fields, it was observed that the Casimir energy is non-vanishing only when the critical exponent associated with space-time anisotropy is an odd number. As a natural extension of this study, we now propose to investigate the Casimir effect for massive fermionic fields in the HL scenario. This approach not only makes the study more relevant for future experimental applications but also introduces greater technical complexity due to the presence of a non-vanishing mass, compared to the analysis developed in \cite{daSilva:2019iwn}.

This paper is structured as follows. In Section \ref{the_model}, we discuss the modified Dirac Lagrangian within the framework of the Horava-Lifshitz approach, considering only the case where the parameter $\xi$, associated with Lorentz symmetry breaking, called the critical exponent, is an odd number. We explicitly present solutions for massive fields confined in the region between two large and parallel plates. Also in this section, we discuss how it is possible to formulate a modified MIT bag model compatible with the higher-order spatial derivative in the Dirac equation. In Section \ref{casimir_eff}, we present the formal expression for the vacuum energy. By using  the generalized Abel-Plana summation formula, we provide an integral expression for renormalized Casimir energy. Considering specific  values of the parameter $\xi$, we present asymptotic results for this observable for large and small values of mass. In addition, we present plots that exhibit the behaviors of the Casimir energy with the mass of the field. The most relevant results are summarized and discussed in section \ref{Concl}. In the Appendices \ref{Ap_a} and \ref{Ap_b}, we provide some important calculations carried out during the development of the work. Here, we adopt natural units $\hbar = c = 1$, and the metric signature convention $(+,-,-,-)$.

\section{The Horava-Lifshitz Model} \label{the_model}

\begin{figure}[h!]
    \centering
    \includegraphics[scale=0.3]{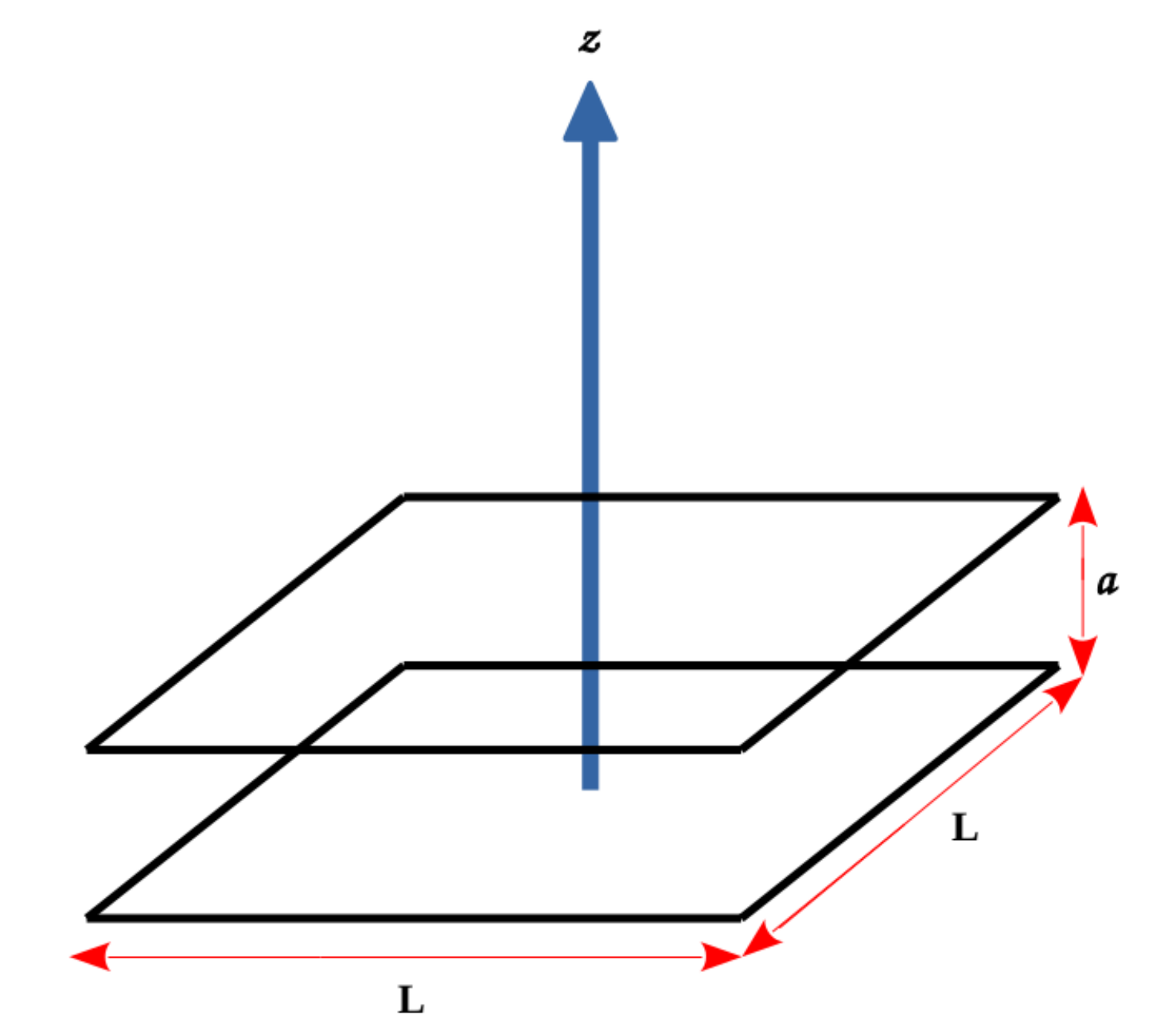}
    \caption{Two large parallel plates with area $L^2$ separated by a distance $a$.}
    \label{fig_plates}
\end{figure}

In this section, we introduce the theoretical model in which the Dirac equation is modified in the context of the Horava-Lifshitz (HL) approach. Our goal is to obtain solutions for this equation that are compatible with the MIT bag model boundary condition, applied to the fields on two parallel plates with areas $L^2$, separated by a distance $a$, being $a \ll L$, as shown in Fig. \ref{fig_plates}.

In the HL context, the Lagrangian density adopted to describe the dynamics of a massive fermionic field, as presented in \cite{Mariz:2019kmc}, is given by:
\begin{eqnarray} \label{Lagran}
    {\cal{L}}={\bar{\psi}}\left\{i\gamma^0\partial_t+l^{\xi-1}(i\gamma^i\partial_i)^\xi- m\right\}\psi. 
\end{eqnarray}
Here, the parameter $l$ has the dimension of length and is introduced to provide the correct dimension for the Lagrangian above. The exponent $\xi$, known as the critical exponent, is associated with the violation of Lorentz symmetry. Note that for $\xi=1$, we recover the usual Dirac Lagrangian without Lorentz violation.

From this point on, we will assume that the critical exponent $\xi$ takes only odd integer values. This choice is based on results obtained in \cite{daSilva:2019iwn}, where it was demonstrated that for even values of $\xi$, the Casimir energy is zero. Thus, the modified Dirac equation takes the following form:
\begin{eqnarray} \label{Dirac_mod}
    {\hat{H}}\psi=\left[-l^{\xi-1}\gamma^0(\nabla^2)^{\frac{\xi-1}{2}}(i\gamma^i\partial_i)+m\gamma^0 \right]\psi=i\frac{\partial \psi}{\partial_t},
\end{eqnarray}
where $\gamma^j$ given below, represents the Dirac matrices: 
\begin{eqnarray} \label{Dirac_matrix}
    \begin{aligned}
	\gamma^0 = \begin{pmatrix}
		I & 0 \\
		0 & -I
	\end{pmatrix} \ \ \ \ \ \ \ \text{and} \ \ \ \ \ \ \ 
	\gamma^j = \begin{pmatrix}
		0 & \sigma^j \\
		-\sigma^j & 0
	\end{pmatrix}  \  . 
    \end{aligned}
\end{eqnarray}
In the definition above, $I$ represents the $2 \times 2$ identity matrix, and $\sigma^j$, $j=1, \ 2, \ 3$ the Pauli matrices.

As mentioned previously, our focus is on calculating the Casimir energy and, consequently, the Casimir pressure associated with the massive fermionic quantum field in the Horava-Lifshitz framework. To achieve this objective, we consider that the fields are confined between two parallel plates as illustrated in Fig. \ref{fig_plates}, considering $L>>a$. Therefore, assuming the time dependence of the fermionic wave function in the form $e^{\mp \omega t}$, we obtain the positive and negative energy solutions for Eq. \eqref{Dirac_mod}, given by:
\begin{eqnarray} \label{func_posit}
    \psi^{(+)}_{\beta}(x) = A_{\beta} e^{-i \omega_{\beta} t} \begin{pmatrix} \varphi^{(+)}(\vec{x}) \\ \chi^{(+)}(\vec{x})\end{pmatrix}  \ \ \ \ \ {\rm with} \ \ \ \ \ \chi^{(+)}(\vec{x})=-\frac{i  l^{\xi-1}(\nabla^2)^{\frac{\xi-1}{2}}}{\omega + m}\sigma^j \partial_j \varphi^{(+)}(\vec{x})
\end{eqnarray}
and
\begin{eqnarray}
	\label{func_negat}
\psi^{(-)}_{\beta}(x) = A_{\beta} e^{i \omega_{\beta} t} \begin{pmatrix} \varphi^{(-)}({\vec{r}}) \\ \chi^{(-)}(\vec{x}) \end{pmatrix}  \ \ \ \ \ {\rm with} \ \ \ \ \ \varphi^{(-)}({\vec{r}})=\frac{il^{\xi-1}(\nabla^2)^{\frac{\xi-1}{2}}}{\omega+m}
\sigma^j\partial_j \chi^{(-)}(\vec{x}), 
\end{eqnarray}
where $\omega_{\beta}$ is the dispersion relation, given by
\begin{eqnarray} \label{Disper}
    \omega_\beta =l^{\xi-1} \sqrt{(\vec{k}^2)^\xi+\mu^{2\xi}}.
\end{eqnarray}
Here, the parameter $\beta$ represents the momentum components ${k_x, k_y, k_z}$. Furthermore, note that in Eq. \eqref{Disper} we have replaced the mass parameter, $m$, with $m = \mu^\xi l^{\xi-1}$.

The Eqs. \eqref{func_posit} and \eqref{func_negat} represent two-component spinors, $\varphi$ and $\chi$, where $\psi_{\beta}^{(+)}$ and $\psi_{\beta}^{(-)}$ correspond to particles and antiparticles, respectively. The components of the spinors have the following explicit form:
\begin{eqnarray} \label{two_comp}
    \begin{aligned}
	\varphi^{(+)}(\vec{x}) = e^{i(k_x x + k_y y)} \Big{(}\varphi_+ e^{ik_z z} + \varphi_- e^{-ik_z z}\Big{)} ,  \ \ \chi^{(-)} (\vec{x}) = e^{-i(k_x x + k_y y)} \Big{(}\chi_+ e^{ik_z z} + \chi_- e^{-ik_z z}\Big{)}.
    \end{aligned}
\end{eqnarray}
As said before, the fermionic modes are characterized by the set of quantum numbers ${k_x, k_y, k_z}$, with $k_x$ and $k_y$ being continuous quantum numbers; however, $k_z$ must be discrete to satisfy the boundary conditions on the plates. Therefore, assuming the standard approach to derive the MIT bag boundary condition, and considering the modified Lagrangian from Eq. \eqref{Lagran}, we obtain:
\begin{eqnarray} \label{MIT_bag_boundary}
	\left(1+il^{\xi-1}(\nabla^2)^{\frac{\xi-1}{2}}\gamma^\mu n_\mu\right)\psi\Big{|}_{z=0, a} =0, 
\end{eqnarray}
with $n_{\mu}$ being the unit vector normal directed outward from the boundaries. For the plate configuration we are considering, we have $n_{\mu} = - \delta_{\mu}^z$ for the plate at $z = 0$ and $n_{\mu} = \delta_{\mu}^z$ at $z = a$.

Now, let's analyze the solutions on the plates to find a transcendental relationship. Considering the positive-energy solution from Eq. \eqref{func_posit}, on the surface $z=a$ with $\vec{n}=-\hat{z}$, the Eq. \eqref{MIT_bag_boundary} provides:
\begin{equation} \label{boundary_1}
	il^{\xi-1}(\nabla^2)^{\frac{\xi-1}{2}}\left(\begin{array}{cc} 0 & \sigma^z \\ -\sigma^z & 0 \end{array}
	\right)\begin{pmatrix} \varphi \\ \chi
	\end{pmatrix}=-\begin{pmatrix} \varphi \\ \chi \end{pmatrix}.
\end{equation}
Considering that $\nabla^2 \varphi = -\vec{k}^2 \varphi$, we have:
\begin{eqnarray} \label{MIT_1}
	\left\{\begin{array}{ll} i^\xi l^{\xi-1}({\vec{k}}^2)^{\frac{\xi-1}{2}}\sigma_3\chi=-\varphi \\ & \\ i^\xi l^{\xi-1}({\vec{k}}^2)^{\frac{\xi-1}{2}}\sigma_3\varphi=\chi . \end{array} \right.
\end{eqnarray}
Now, substituting $\chi(\vec{x})$ given in \eqref{func_posit} into \eqref{MIT_1}, we notice that the second equation is not satisfied for arbitrary values of the critical exponent $\xi \neq 1$. However, if we are inclined to accept that the modified Dirac equation can confine electrons in a finite region, we must appropriately modify the MIT bag model so that for $\xi = 1$, we recover the standard MIT model. In Appendix \ref{Ap_a} we propose a  higher-order derivative Lagrangian associated with the MIT bag model that solves this problem. Accepting this modified model, the equations relating the two-component spinors $\varphi$ and $\chi$ on the two flat boundaries, coincide with those corresponding to $\xi = 1$, as shown in Eqs. \eqref{cond_1} and \eqref{cond_2} on the plate $z=a$, for positive and negative energy solutions, respectively. They are given by:
\begin{eqnarray} \label{cond_a}
    \sigma^3\varphi^{(+)}= \frac{\sigma^i\partial_i\varphi^{(+)}}{w+m} \ \ \ \ \ \text{and} \ \ \ \ \ 
    \sigma^3\chi^{(-)}=-\frac{\sigma^i\partial_i\chi^{(-)}}{w+m} .
\end{eqnarray}
From the boundary condition on the plate at $z = 0$, we obtain the following relations between the spinors given in \eqref{two_comp}:
\begin{eqnarray}
	\label{timelike_spinor_1}
	\begin{aligned}
		\varphi_+ = - \frac{m(\omega + m) + k_z^2 - k_z \sigma^3(\sigma^1 k_x+\sigma^2 k_y)}{(m-ik_z)(\omega + m)} \varphi_- , \\
		\chi_- = - \frac{m(\omega+m)+k_z^2-k_z\sigma^3(\sigma^1k_x+\sigma^2k_y)}{(m+ik_z)(\omega+m)} \chi_+ .
	\end{aligned}
\end{eqnarray}
Finally, considering the boundary condition for the fermionic functions at $z=a$, we arrive at the following transcendental equation:
\begin{eqnarray} 
	\label{trans_equ}
    \frac{a m}{k_z a} \sin(k_z a) + \cos(k_z a) = 0  \   .
\end{eqnarray}
We just want to emphasize that the above equation coincides with the standard equation for the Lorentz preserving Dirac equation. 

\section{The influence of Lorentz symmetry breaking in Casimir effect} \label{casimir_eff}

In this section, we will focus on calculating the Casimir energy associated with the quantum fermionic vacuum fluctuations considering  the Horava-Lifshitz Lorentz violation scenario. To this objective, we adopt the formalism described in \cite{Itzykson:1980rh} and present the explicit expansion of the field operator in terms of the normalized positive and negative energy fermionic modes: 
\begin{equation} \label{field_operator}
    \psi (x) = \sum_{r, n} \int dk_x \int dk_y \left[ a_{r,\beta} \psi^{(+)}_{r,\beta} + b_{r,\beta}^{\dag}\psi^{(-)}_{r,\beta}\right],	
\end{equation}
where $a_{r, \beta}$ and $b_{r, \beta}^{\dag}$ are the annihilation and creation operators for particles and antiparticles, respectively, with momentum $\vec{k}$ and polarization $r$.

The standard Dirac Lagrangian provides the following Hamiltonian density operator, as demonstrated in \cite{Mandl:1985bg}:
\begin{equation} \label{hamilt_ope}
    \mathcal{H} = i \psi^{\dag} \dot{\psi} \ .
\end{equation}
Following a similar procedure, we verify that the modified Dirac Lagrangian in Eq. \eqref{Lagran} yields the same expression for the Hamiltonian density.

Therefore, the vacuum energy, $E_0$, is determined by calculating the expected value of the Hamiltonian operator from \eqref{hamilt_ope} in the vacuum state, as shown below:
\begin{equation} \label{vac_ener}
    E_0 = i \int_V d^3x \langle 0 | \psi^{\dag} \dot{\psi} | 0 \rangle \ .
\end{equation}

Substituting the field operator given in \eqref{field_operator} into \eqref{vac_ener}, we obtain:
\begin{equation} \label{Vacuum_energy}
    E_0 =  \sum_{r}\sum_{n}\int  d^2{\vec{k}} \ \omega_{\beta} \langle 0|a_{r,\beta}^\dagger a_{r,\beta} - b_{r,\beta} b_{r,\beta}^{\dag} |0 \rangle   \  .
\end{equation}
Furthermore, by using the standard anti-commutation relations for the operators $a_{r,\beta}$ and $b_{r,\beta}$, the above expression provides
\begin{equation} \label{vac_ene_2}
    E_0 = - \frac{L^2  l^{\xi - 1}}{4 \pi^2} \sum_{n} \int  d^2{\vec{k}} \sqrt{(\vec{k}^2)^\xi+\mu^{2\xi}}  \  .
\end{equation}
Here, we have $\vec{k}^2 = k_x^2 + k_y^2 + k_z^2$. Expressing $k_x$ and $k_y$ in polar coordinates, the above expression reads,
\begin{equation} 
	\label{Vacuum_energy1}
    E_0 = - \frac{L^2 l^{\xi - 1}}{\pi} \int_0^{\infty} k dk \sum_{n=1}^{\infty} \sqrt{\left[k^2+\left({\alpha_n}/{a}\right)^2\right]^{\xi}+\mu^{2\xi}} \ ,
\end{equation}
where we have defined $\alpha_n=k_na$, with $k_n=k_z$ being the $n$-th root of the transcendental equation given in \eqref{trans_equ}. As we can see, the vacuum energy above is divergent. Therefore, in order to obtain a finite and and well-defined result, namely the Casimir energy, we have to renormalize this expression. This can be done by employing the Abel-Plana summation formula \cite{Saharian:2006iv}:
\begin{eqnarray}
	\label{abel_plana}
	\sum_{n=1}^{\infty} \frac{\pi f(\alpha_n)}{1-\frac{\sin(2\alpha_n)}{2\alpha_n}} = - \frac{\pi am f(0)}{2(am+1)} + \int_0^{\infty} dz f(z) - i\int_0^{\infty} dt \frac{f(it)-f(-it)}{\frac{t+am}{t-am}e^{2t}+1} \ .
\end{eqnarray}

To develop the summation over $n$ in \eqref{Vacuum_energy1} we  must rewrite the integrand in an appropriate manner. At this point we follow the steps adopted in \cite{Aram_2009}. So let  us start with the denominator appearing in the left hand side of the summation \eqref{abel_plana}. By using \eqref{trans_equ}, it is possible to obtain the identity below:
\begin{eqnarray}
    1-\frac{\sin(2\alpha_n)}{2\alpha_n} = 1 + \frac{am}{(am)^2+\alpha_n^2}   \ .
\end{eqnarray}
The function $f(x)$ in Eq. \eqref{abel_plana} is defined as
\begin{eqnarray} \label{func_f}
    f(x) = \sqrt{(x^2+k^2a^2)^{\xi}+\mu^{2\xi}a^{2\xi}}\left(1+\frac{am}{(am)^2+x^2}\right).
\end{eqnarray}
Thus, substituting \eqref{abel_plana} into \eqref{Vacuum_energy1}, the vacuum energy becomes,
\begin{eqnarray} \label{vac_ene_1}
    \begin{aligned}
    E_0 = - \frac{L^2 l^{\xi-1}}{\pi^2 a^{\xi}} \int_0^{\infty}kdk &\Bigg \{ - \frac{\pi am}{2(am+1)}f(0) + \int_{0}^{\infty}f(z)dz - i \int_{0}^{\infty} \frac{f(it)-f(-it)}{\frac{t+am}{t-am}e^{2t}+1}dt \Bigg\}  \ .
    \end{aligned}
\end{eqnarray}
Note that the equation above involves the mass $m$ and also the quantity $\mu$, which is related to the mass. Although being possible to use just one of  them, here we will keep both notations. In the final expressions, we will express $\mu$ in terms of the mass $m$.

The first term inside the brackets in \eqref{vac_ene_1} corresponds to the vacuum energy in the presence of only one plate, while the second term is related to the vacuum energy in the absence of plates. Therefore, these terms are divergent, but after renormalization, they are discarded. Consequently, the Casimir energy is associated with the third term, and is given by:
\begin{eqnarray}
    E_C &=& \frac{iL^2l^{\xi-1}a}{\pi^2} \int_0^{\infty}kdk \int_0^{\infty}  \frac{u-m}{(u+m)e^{2au}+u-m}\left(1+\frac{m}{am^2-au^2}\right) \nonumber\\
     & \times& \left \{ \sqrt{[(iu)^2+k^2]^{\xi}+\mu^{2\xi}} - \sqrt{[(-iu)^2+k^2]^{\xi}+\mu^{2\xi}}\right \} du  \  ,
\end{eqnarray}
where we have performed the substitution $t=au$. At this point, we should divide the integral with respect to the variable $u$ into two intervals: the first is $\left[0, \sqrt{k^2+\mu^2}\right]$ and the second is $\left[\sqrt{k^2+\mu^2}, \infty \right)$. However, we find that the integral in the first interval vanishes, and, consequently, the resulting Casimir energy is given by,
\begin{eqnarray} 
	\label{cas_energ}
\begin{aligned}
    E_C = - \frac{2L^2l^{\xi-1}}{\pi^2} \sin{\left(\frac{\pi \xi}{2}\right)} \int_0^{\infty} kdk \int_{\sqrt{k^2+\mu^2}}^{\infty} \frac{\left[a(u-m)-\frac{m}{u+m}\right] \sqrt{(u^2-k^2)^{\xi}-\mu^{2\xi}}}{(u+m)e^{2au}+u-m} du \  .
\end{aligned}
\end{eqnarray}

In order to obtain a more suitable expression that allows us to analyze asymptotic limits for the dimensionless quantity 
$am$, we will adopt the following change of coordinates (see Appendix \ref{Ap_b}), $x=\sqrt{u^2-k^2-\mu^2}$. Expressing  $x=r\cos\theta$ and $k=r\sin\theta$,  the Casimir energy can be written as:
\begin{eqnarray} \label{Energy_1}
    E_C = - \frac{mL^2}{\pi^2} \sin{\left(\frac{\pi \xi}{2}\right)} \int_0^{\infty} r^3 \frac{g(\sqrt{r^2+\mu^2})}{\sqrt{r^2+\mu^2}} dr \int_0^1 \sqrt{\left({r^2v}/{\mu^2}+1\right)^{\xi}-1} \ dv \ ,
\end{eqnarray}
where
\begin{eqnarray}
    g(y) =\frac{a(y-m)-\frac{m}{y+m}}{(y+m)e^{2ay}+y-m} \  .
\end{eqnarray}
Finally, adopting the variable change $z=ar$, we obtain
\begin{eqnarray} 
	\label{Energy_2}
\begin{aligned}
    E_C = - \frac{L^2 m}{\pi^2 a^2} \sin{\left(\frac{\pi \xi}{2}\right)} & \int_0^{\infty}  {\frac{\sqrt{{\mu}^{2}{a}^{2}+{z}^{2}}-ma-{\frac{ma}{\sqrt{{\mu}^{2}{a}^{2}+{z}^{2}}+ma}}}{\left(\sqrt{{\mu}^{2}{a}^{2}+{z}^{2}}+ma \right){e^{2 \sqrt{{\mu}^{2}{a}^{2}+{z}^{2}}}}+\sqrt{\mu^{2}a^{2}+{z}^{2}}-ma}} \\ & 
    \times \frac{z^3 dz}{\sqrt{z^2+\mu^2 a^2}} \int_0^1 \sqrt{\left({z^2v}/{(\mu a)^2}+1\right)^{\xi}-1} \ dv \ .
    \end{aligned}
\end{eqnarray}
As mentioned earlier, the final results will be expressed in terms of the mass $m$. For this, we consider $a \mu =(am)^{1/\xi} \rho_{\xi}$, where $\rho_{\xi}=(a/l)^{1-1/\xi}$. Because the parameter $l$ is of the order of the inverse of Planck energy scale, the parameter $\rho_{\xi}$ is much greater than unity.

\begin{figure}[h!]
    \centering
    \begin{subfigure}[b]{\textwidth}
        \centering
        \includegraphics[width=0.6\textwidth]{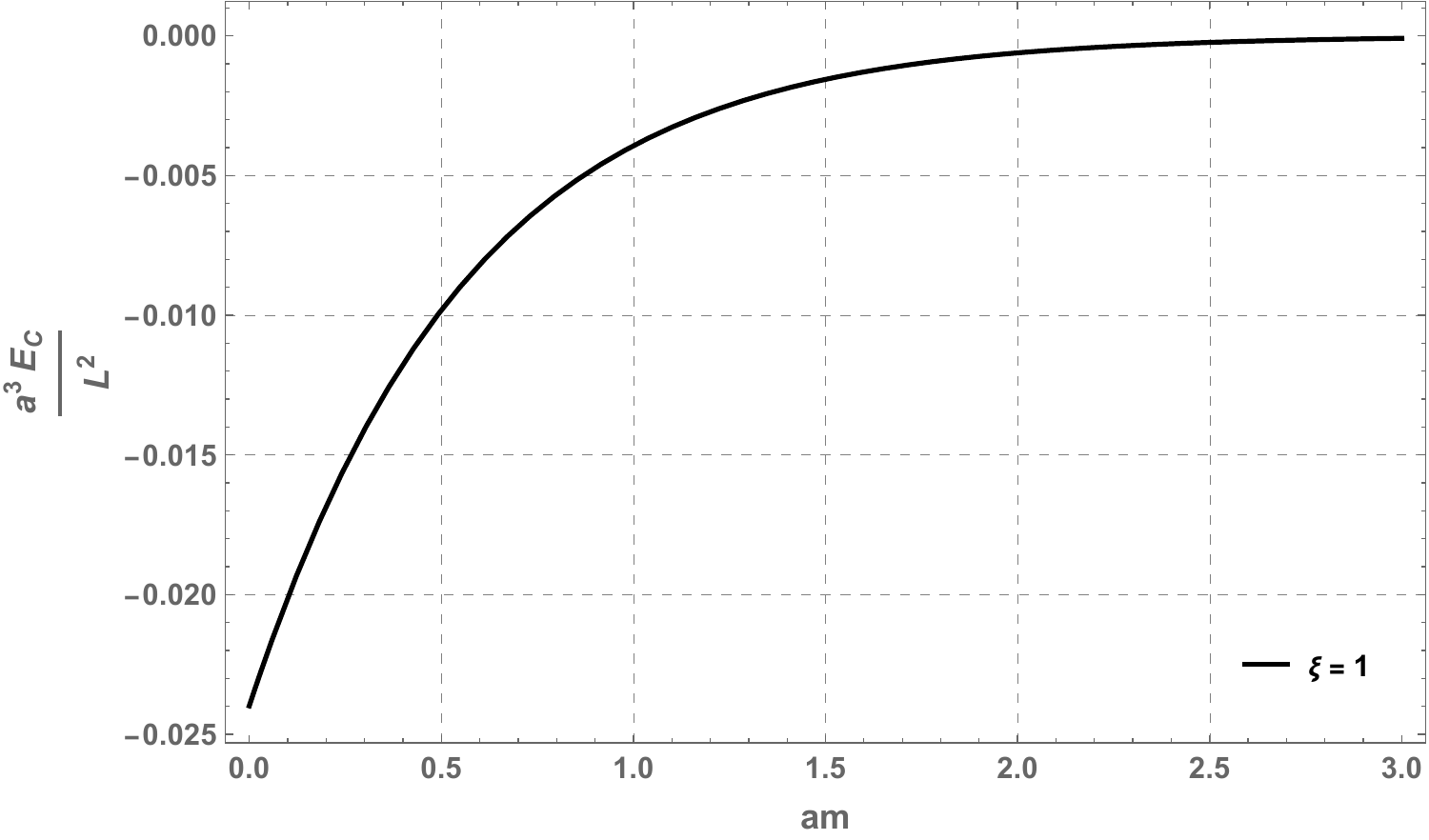}
        \caption{For $\xi=1$.}
    \end{subfigure}
    \vfill
    \begin{subfigure}[b]{\textwidth}
        \centering
        \includegraphics[width=0.6\textwidth]{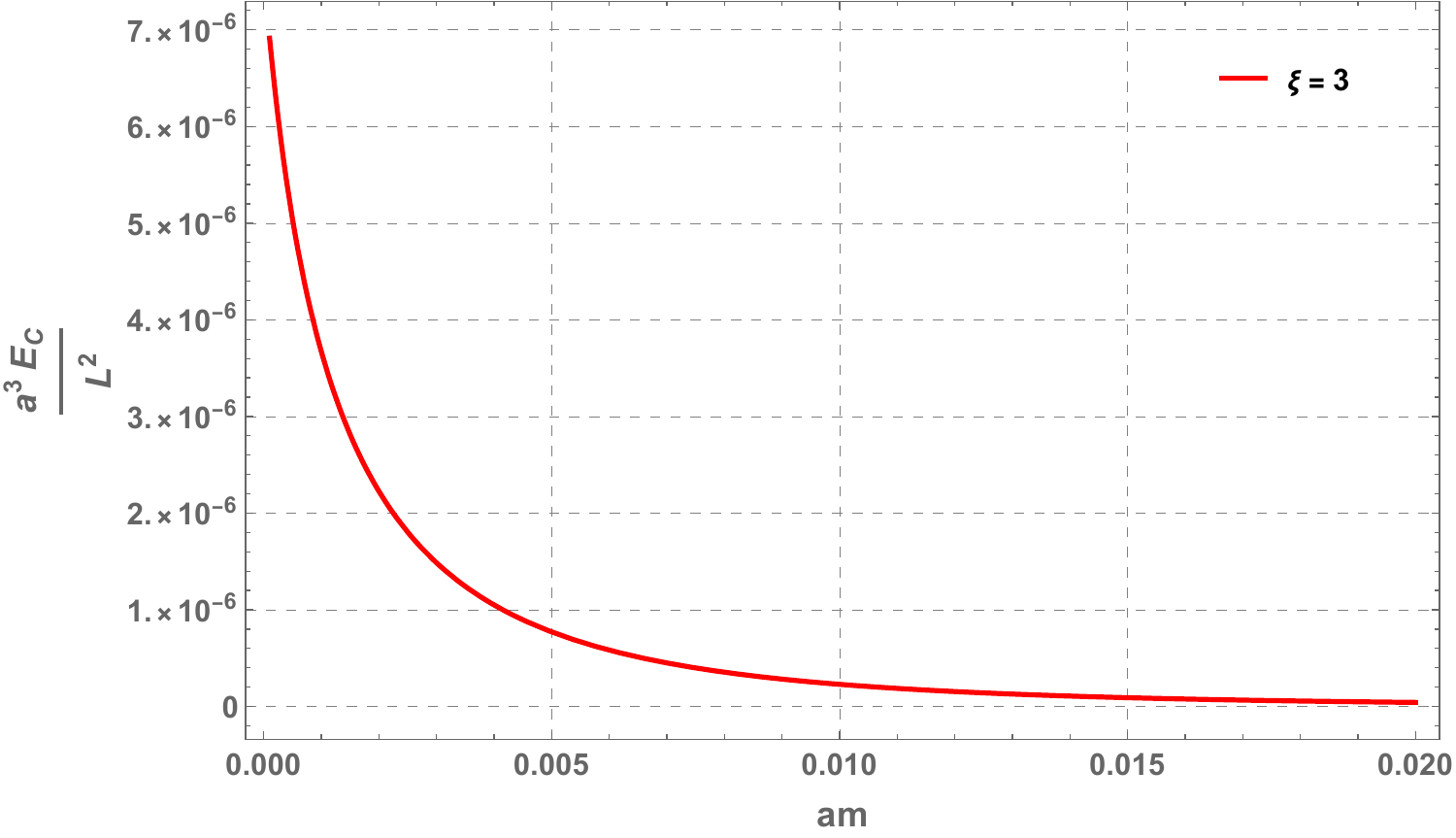}
        \caption{For $\xi=3$.}
    \end{subfigure}
    \vfill
    \begin{subfigure}[b]{\textwidth}
        \centering
        \includegraphics[width=0.6\textwidth]{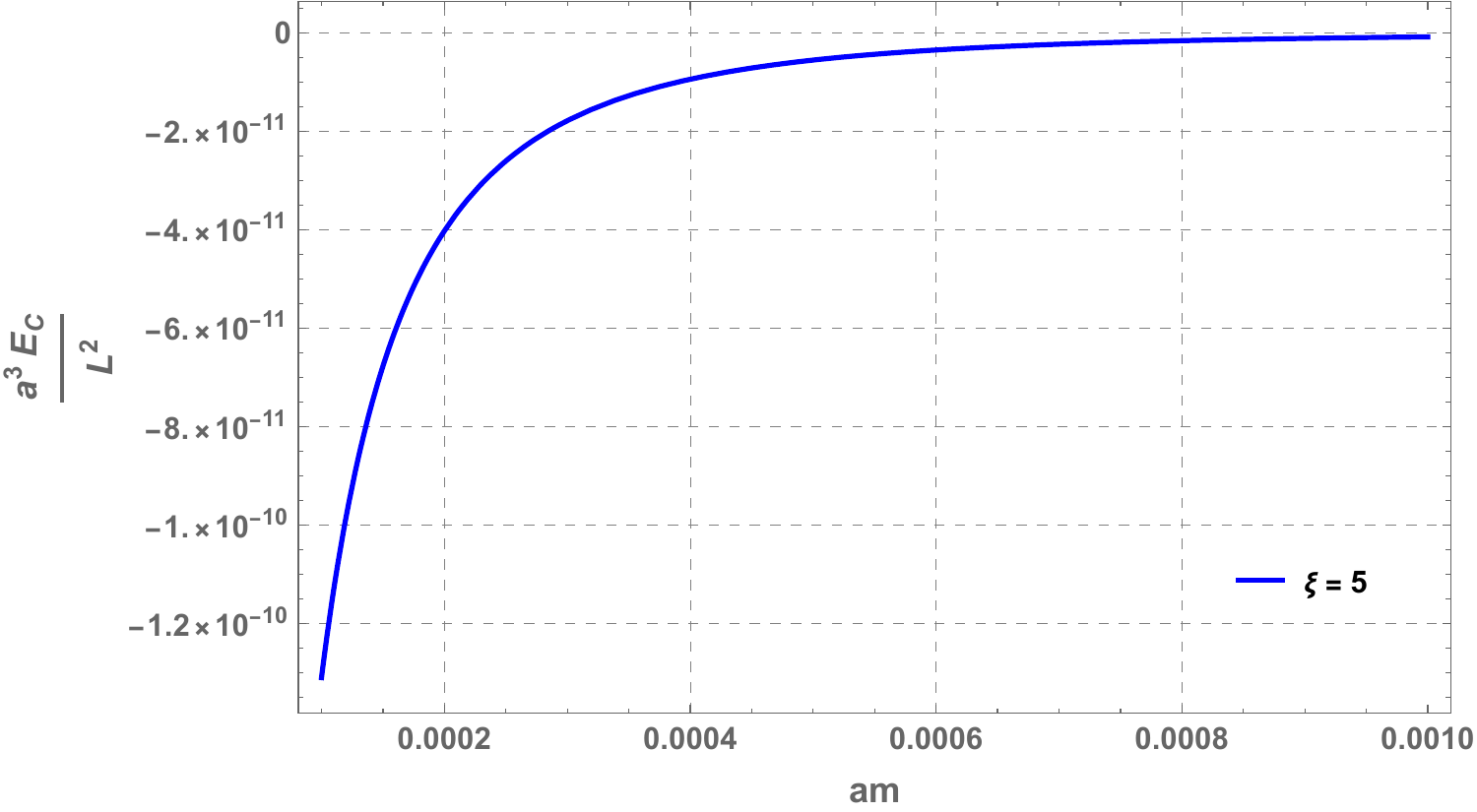}
        \caption{For $\xi=5$.}
    \end{subfigure}
        \caption{The behavior of the Casimir energy, Eq. \eqref{Energy_2}, in units of $L^2/a^3$, as function of $a m$, considering different values of $\xi$, as $\xi=1$ (a), $\xi=3$ (b), and $\xi=5$ (c). Thus, Lorentz symmetry violation affects the Casimir energy both in intensity and in sign.}
    \label{plot_casimir}
\end{figure}

Unfortunately, for $m\neq0$ it is not possible to perform the integral in \eqref{Energy_2}, even for $\xi=1$. Note that, for $\xi=1$, we recover the Casimir energy without the influence of Lorentz symmetry violation. In this case, the result is consistent with the Casimir energy obtained in \cite{Cruz:2018thz} when Lorentz violation is neglected. 
Moreover,  for $\xi>1$, Eq. \eqref{Energy_2} presents an additional difficulty, that is an extra integral in the variable $v$. However, for massless fermionic fields the integral in \eqref{Energy_2} can be exactly solved, and the result agrees with the result found in \cite{daSilva:2019iwn}. Therefore, to provide a better understanding about the behavior of Casimir energy for massive fields with Lorentz violation, we present in Fig. \ref{plot_casimir} three plots of the Casimir energy for $\xi=1, \ 3 , \ 5$. For these three graphs, we consider only as an illustrative example $\frac {a}{l}=10^2$. From them we can observe a change in sign and intensity that is strongly dependent on $\xi$. For $\xi=1, \ 5$ the forces between the plates are attractive, being repulsive for $\xi=3$, and the decay becomes more pronounced with increasing of $\xi$.

As mentioned earlier, we cannot obtain a closed expression for the integrals in Eq. \eqref{Energy_2}, i.e., to express it in terms of spacial functions. However, by this result it is possible to explore some important  behaviors. These limiting cases can be interpreted as corresponding to small mass or distances between plates,  and large mas or distances  between them, respectively. Let's analyze each case separately:
\begin{itemize}
    \item However, before to develop these behaviors, let us obtain the Casimir energy for the case of massless fermions. After some intermediate steps, from \eqref{Energy_2}, we get:
    \begin{eqnarray}
        \begin{aligned}
            E_C = - \frac{L^2 l^{\xi-1} 2^{-2(\xi +2)} \left(2^{\xi+2}-1\right) \Gamma(2+\xi) \zeta(3+\xi) \sin{(\pi \xi /2)}}{\pi^2 a^{\xi +2}} \ .
        \end{aligned}
    \end{eqnarray}
    Consequently, the Casimir pressure is expressed as:
    \begin{eqnarray}
    \begin{aligned}
            P_C = \frac{F_C}{L^2} = - \frac{l^{\xi-1} 2^{-2(\xi +2)} \left(2^{\xi+2}-1\right) \Gamma(3+\xi) \zeta(3+\xi) \sin{(\pi \xi /2)}}{\pi^2 a^{\xi+3}} \  .
        \end{aligned}
    \end{eqnarray}
Both expressions above coincide with obtained previously in \cite{daSilva:2019iwn}.
\end{itemize}

Now, let's consider the limiting cases when $am \ll 1$ and $am \gg 1$. For $am<<1$, the obtained result for any values of $\xi$, is a long expression and not very enlightening. Therefore, we consider only the cases exhibited in Fig. \ref{plot_casimir}, namely, the cases $\xi=1, 3, 5$. 
\begin{itemize}
    \item For the case where $am \ll 1$, we have:
    \begin{itemize}
        \item For $\xi=1$, the integral in $v$ in Eq. \eqref{Energy_2} gives us:
            \begin{eqnarray} 
            	\label{int_v1}
            \begin{aligned}
                \int_0^1 \sqrt{\left({z^2v}/{(\mu a)^2}+1\right)^{\xi}-1} \ dv = \frac{2z}{3 a m} .
            \end{aligned}
            \end{eqnarray}
        Now, substituting this result into \eqref{Energy_2}, the integral over $z$ can be approximated evaluated providing for the Casimir energy the approximated result:
    \begin{eqnarray}
        \begin{aligned}
        	\label{Energy_approx_1}
            E_C \approx - \frac{L^2}{\pi^2 a^3} \left(\frac{7 \pi^4}{2880}-\frac{\pi^2 a m}{24}\right) \ .
        \end{aligned}
    \end{eqnarray}
    As to the the Casimir pressure is get:
    \begin{eqnarray}
    	\label{Pressure_approx_1}
        \begin{aligned}
            P_C \approx - \frac{7\pi^2}{960 a^4}\left(1-\frac{80 a m}{7\pi^2}\right) .
        \end{aligned}
    \end{eqnarray}
    As expected, this result agrees with that obtained in \cite{Cruz:2018thz} for the case where Lorentz symmetry is not violated.

    \item For $\xi=3$, the integral in $v$ is
    \begin{eqnarray} 
    	\label{int_v3}
            \begin{aligned}
                \int_0^1 & \sqrt{\left({z^2v}/{(\mu a)^2}+1\right)^{\xi}-1} \ dv = \frac{2 \left(a^2\mu^2+z^2\right) \sqrt{3 a^4\mu^4+3 a^2\mu^2 z^2+z^4}}{5 a^3\mu^3 z} \\ & + \frac{3^{3/4} a^2\mu^2 F\left(\frac{2 \sqrt[4]{3} a\mu z}{\sqrt{3} a^2\mu^2+z^2}, \frac{1}{\sqrt{2} \left(1+\sqrt{3}\right)}\right)}{5 z^2} -\frac{2\ 3^{3/4} a^2\mu^2 K\left[\frac{1}{\sqrt{2}\left(1+\sqrt{3}\right)}\right]}{5 z^2} \ .
            \end{aligned}
    \end{eqnarray}
    Where the special functions $F$ and $K$ represent the elliptic integral of the first and second kinds, respectively \cite{Abramowitz}. Thus, substituting the above expression into \eqref{Energy_2}, and considering $a\mu=\sqrt[3]{ma} (a/l)^{2/3}$, the Casimir energy reads,
    \begin{eqnarray}
        \begin{aligned}
        		\label{Energy_approx_3}
            E_C \approx \frac{L^2}{\pi ^2 a^3} \left(\frac{l}{a}\right)^2 \left(\frac{31 \pi^6}{40320}-\frac{7 \pi^4 am}{960}\right) .
        \end{aligned}
    \end{eqnarray}
    Therefore, the Casimir pressure is
    \begin{eqnarray}
  \label{Pressure_approx_3}  	
        \begin{aligned}
            P_C \approx \left(\frac{l}{a}\right)^2 \left(\frac{31 \pi^4}{8064 a^4}-\frac{7 \pi^2 a m}{240 a^4}\right) .
        \end{aligned}
    \end{eqnarray}
    \item Finally, for $\xi=5$, the integral in $v$ is given by
    \begin{eqnarray} 
    	\label{int_v5}
                &&\int_0^1  \sqrt{\left({z^2v}/{(\mu a)^2}+1\right)^{\xi}-1} = - \frac{(\sqrt{5}-1) \pi^{3/2} a^2\mu^2}{7 z^2 \Gamma(7/10) \Gamma(4/5)}\nonumber\\
                & +& \frac{2 \left(a^2\mu^2+z^2\right)^{7/2} \,
                _2F_1\left(-\frac{7}{10}, -\frac{1}{2}; \frac{3}{10}; \frac{a^{10}\mu^{10}}{\left(a^2\mu^2+z^2\right)^5}\right)}{7 a^5\mu^5 z^2} \ .
    \end{eqnarray}
In the above expression, $ _2F_1$ represents the hypergeometric function \cite{Abramowitz}.

For this case, an approximated expression can be obtained for the integral with respect to $z$ in \eqref{Energy_2},  consequently, in this approximation, the Casimir energy is expressed by
    \begin{eqnarray} 
    \label{Energy_approx_5}	
        \begin{aligned}
            E_C \approx - \frac{L^2}{\pi^2 a^3} \left(\frac{l}{a} \right)^4 \left(\frac{127 \pi^8}{215040} - \frac{31 \pi^6 a m}{8064}\right) \ ,
        \end{aligned}
    \end{eqnarray}
    and the pressure by
    \begin{eqnarray}
    \label{Pressure_approx_5}	 
        \begin{aligned}
            P_C \approx - \left(\frac{l}{a}\right)^4 \left(\frac{127 \pi^6}{30720 a^4} - \frac{31 \pi^4 a m}{1344 a^4}\right) \ .
        \end{aligned}
    \end{eqnarray}
\end{itemize}
    Thus, we can observe that the critical exponent significantly impacts the Casimir energy and, consequently, the Casimir pressure. In the case where $\xi=1$, i.e., the scenario without Lorentz violation, the standard result is obtained, as expected. However, the new results for $\xi>1$ can provide a fresh perspective on the study of potential Lorentz violation.  

    \item Now, for the case where $am \gg 1$, the integral with respect to $v$ can be approximated by:
    \begin{eqnarray} \label{int_vxi}
            \begin{aligned}
                \int_0^1  \sqrt{\left({z^2v}/{(\mu a)^2}+1\right)^{\xi}-1} & \ dv \approx \frac{2 \sqrt{\xi} z}{3 a \mu} \ .
            \end{aligned}
    \end{eqnarray}
    Thus, the integration with respect to $z$ in \eqref{Energy_2} yields:
    \begin{eqnarray}
        \begin{aligned}
        I_{\xi} \approx (am)^{\frac{1}{\xi}} (a/l)^{1-\frac{1}{\xi}} e^{-2 (am)^{\frac{1}{\xi}} (a/l)^{1-\frac{1}{\xi}}} .
        \end{aligned}
    \end{eqnarray}
Substituting this result back into \eqref{Energy_2}, we get
\begin{eqnarray}
    \begin{aligned}
    	\label{Energy_L}
        E_C \approx - \frac{L^2 (am)^{\frac{1}{\xi}} (a/l)^{1-\frac{1}{\xi}} \sin{(\pi \xi/2)}}{\pi^2 a^3} e^{-2 (am)^{\frac{1}{\xi}} (a/l)^{1-\frac{1}{\xi}}} .
    \end{aligned}
\end{eqnarray}
Here we can see that the Casimir energy decays exponentially with $(am)^{\frac{1}{\xi}} (a/l)^{1-\frac{1}{\xi}}$. Therefore, in this limit, the Casimir pressure also quickly approaches zero. This result is consistent with the behavior of the graphs in Fig. \ref{plot_casimir}. Additionally, we can note that in this limit, calculations can proceed for any odd $\xi$.
\end{itemize}

\section{Concluding Remarks} \label{Concl}
In this paper, we analyze the fermionic Casimir effects within the HL formalism of Lorentz violation, considering that a massive fermionic field is confined in the region between two large parallel plates. To implement this confinement, it was necessary to modify the Lagrangian density associated with the MIT bag model to make it compatible with the higher-order spatial derivative system. This modified model is explicitly presented in Eq. \eqref{Lagrang} of Appendix \ref{Ap_a}. We then found the transcendental equation \eqref{trans_equ}, which discretizes the momentum in the direction orthogonal to the plates. Using the generalized Abel-Plana summation formula, Eq. \eqref{abel_plana}, we derived an expression for the Casimir energy in \eqref{Energy_2}, based on the general vacuum energy given in Eq. \eqref{Vacuum_energy1}.

Indeed, in our analysis, we considered only cases where the critical exponent, $\xi$, is an odd number. For even values of $\xi$, the corresponding Hamiltonian operator becomes diagonal, and the spin-up and spin-down components of the fermionic field do not interact with each other. Just like in the case of $\xi=1$, the result for the Casimir energy, Eq. \eqref{Energy_2}, is given in terms of combinations of special functions. Furthermore, for $\xi>1$, this expression presents an additional difficulty represented by an extra integral in the variable $v$. In principle, it is possible to find a general result for this integral, but the expression is long and not very illuminating. Therefore, given this fact, we decided to present results for this integral only for $\xi=1, \ 3 , \ 5$, as given in \eqref{int_v1}, \eqref{int_v3}, and \eqref{int_v5}, respectively. To provide a better understanding of the behavior of the fermionic Casimir energy, we present in Fig. \ref{plot_casimir} three plots showing the Casimir energy for $\xi=1, \ 3 , \ 5$ as a function of the dimensionless parameter $am$. For these three graphs, we considered, as an illustrative example, $\frac {a}{l}=10^2$. As we can observe, the corresponding energies exhibit a strong dependence on the critical exponent, changing their signs and intensities.
 
Another analysis developed in this paper concerns the behavior of the Casimir energy and pressure in the limiting cases of the dimensionless parameter $am<<1$ and $am>>1$. For $am<<1$, we present the asymptotic expressions for the Casimir energy and pressure, considering $\xi=1, \ 3, \ 5$. These expressions are given by \eqref{Energy_approx_1} and \eqref{Pressure_approx_1}, \eqref{Energy_approx_3} and \eqref{Pressure_approx_3}, and finally \eqref{Energy_approx_5} and \eqref{Pressure_approx_5}, respectively. In the limit of large $am$, we were able to present the behavior of the Casimir energy in a compact form for any odd value of $\xi$. This expression is given in \eqref{Energy_L}, where we observe a strong exponential decay, accentuated when $\xi>1$.

Therefore, we hope that the results presented here can guide theoretical and experimental studies in exploring aspects related to Lorentz symmetry violation. This will enable us to further improve our theoretical models, including the one in this current work, as well as our experimental framework. We would like to finish this paper by saying that in previous work \cite{Cruz:2018thz}, we have analyzed the Casimir energy in a CPT-even, aether-like Lorentz symmetry violation. In this context the Lorentz symmetry is broken in the Planck scale energy, and this violation is implemented by the emergence of a nonzero vacuum expectation value of some vector and tensor 
components, which implies in preferential directions, therefore, space-time anisotropy \cite{Kostelecky:1988zi}.
Becasue in this present analysis we have analyzed the fermionic Casimir effects considering HL Lorentz violation scenario, we hope to have filled the gap left.
\section*{Acknowledgments}
We would like to thank Aram A. Saharian for  helpful discussions.
\appendix
\section{MIT bag model in the Horava-Lifshitz scenario}
\label{Ap_a}
In this appendix, we introduce a modified Lagrangian in the Horava-Lifshitz Lorentz violation scenario for the MIT bag model, which provides consistent equations for the two-component spinors $\varphi$ and $\chi$. We impose that the positive and negative energy solutions, given in Eqs. \eqref{func_posit} and \eqref{func_negat}, satisfy the MIT boundary condition on the plates located at $z=0$ and $z=a$, as shown in Fig. \ref{fig_plates}. The proposed Lagrangian density given below, is very similar with the standard MIT model:
\begin{eqnarray}
	\label{Lagrang}
		\mathcal{L} &=& \Big{\{} \frac{i}{2}
		\Big{[}\bar{\Psi} \gamma^{0} (\partial_{0}\Psi) - (\partial_{0}\bar{\Psi})\gamma^{0}\Psi\Big{]} +\frac{il^{\xi-1}}{2}\Big{[}\bar{\Psi}(\nabla^2)^{\frac{\xi-1}{2}} \gamma^{i} (\partial_{i}\Psi) - (\partial_{i}\bar{\Psi})(\nabla^2)^{\frac{\xi-1}{2}}\gamma^{i}\Psi\Big{]}\nonumber\\ &-&m\bar{\Psi}\Psi - B \Big{\}}\theta_v- \frac{1}{2} \Delta_s\bar{\Psi}M\Psi \ ,
\end{eqnarray}
where $M$ is a specific $4 \times 4$ matrix operator, as indicated below,
\begin{eqnarray}
    \begin{aligned}
        M= \begin{pmatrix} {\hat{\alpha}} & 0 \\ 0 & {\hat{\beta}}
	\end{pmatrix} .
    \end{aligned}
\end{eqnarray}
Here, ${\hat{\alpha}}$ and ${\hat{\beta}}$ are higher-order differential operators for $\xi > 1$. Both operators reduce to the $2 \times 2$ identity matrix when $\xi = 1$. Furthermore, for positive-energy solutions, we impose ${\hat{\alpha}} = l^{2(\xi-1)}(\nabla^2)^{\xi-1}$ and ${\hat{\beta}} = 1$, and for negative-energy solutions, we impose ${\hat{\alpha}} = 1$ and ${\hat{\beta}} = l^{2(\xi-1)}(\nabla^2)^{\xi-1}$.

In Eq. \eqref{Lagrang}, $\theta_v$ is the Heaviside function, which is equal to unity inside the bag and vanishes outside it. The factor $B$ is related to the energy-momentum tensor on the bag, and $\Delta_s$ is the Dirac delta function given by the spatial derivative of $\theta_v$, as shown below,
\begin{eqnarray}
    \frac{\partial\theta_v}{\partial x^i}=n_i\Delta_s \ , 
\end{eqnarray}
where $n_i$ is the unit vector normal to the surface.

Therefore, using the usual Euler-Lagrange formalism, the dynamics of the system are governed by the modified Dirac equation:
\begin{eqnarray}
    \frac{\partial \mathcal{L}}{\partial \bar{\Psi}} - \partial_{0} \Big{[} \frac{\partial \mathcal{L}}{\partial (\partial_{0} \bar{\Psi})} \Big{]} - \partial_{i} \Big{[} \frac{\partial \mathcal{L}}{\partial (\partial_{i} \bar{\Psi})} \Big{]} =0  \ .
\end{eqnarray}
Therefore, substituting Eq. \eqref{Lagrang} into the equation above, we obtain
\begin{eqnarray}
	\label{EM1}
	\left[i\gamma^{0}(\partial_{0}\Psi)+il^{\xi-1}(\nabla^2)^{\frac{\xi-1}{2}}\gamma^i\partial_i \Psi - m\Psi\right] \theta_v -\frac i2\left[l^{\xi-1}(\nabla^2)^{\frac{\xi-1}{2}}\gamma^\mu n_\mu \Psi -iM\Psi\right] \Delta_s = 0 .
\end{eqnarray}

From Eq. \eqref{EM1}, we can derive two different equations: (i) In the region inside the bag, where $\Delta_s=0$ and $\theta_v=1$, we have
\begin{eqnarray}
    i \gamma^{0}(\partial_{0}\Psi)+i l^{\xi-1}(\nabla^2)^{\frac{\xi-1}{2}}\gamma^i\partial_i \Psi - m\Psi = 0 .
\end{eqnarray}
(ii) On the surface of the bag, where $\Delta_s=\infty$ and $\theta_v=0$, we obtain
\begin{eqnarray}
    l^{\xi-1}(\nabla^2)^{\frac{\xi-1}{2}}\gamma^\mu n_\mu \Psi =iM\Psi ,
\end{eqnarray}
\begin{eqnarray} \label{MIT_2}
    \left\{ 
    \begin{array}{ll}
    i^\xi l^{\xi-1}({\vec{k}}^2)^{\frac{\xi-1}{2}}\sigma_3\chi=-{\hat{\alpha}}\varphi \\ & \\  i^\xi l^{\xi-1}({\vec{k}}^2)^{\frac{\xi-1}{2}}\sigma_3\varphi={\hat{\beta}}\chi \ .
    \end{array} \right.
\end{eqnarray}

Therefore, considering the positive and negative-energy solutions on the plate at $z=a$, both equations above are consistent, resulting:
\begin{eqnarray} \label{cond_1}
    \sigma^3\varphi^{(+)}=-\frac{\sigma^i\partial_i\varphi^{(+)}}{w+m} \ , 
\end{eqnarray}
and 
\begin{eqnarray} \label{cond_2}
    \sigma^3\chi^{(-)}=-\frac{\sigma^i\partial_i\chi^{(-)}}{w+m} \ .
\end{eqnarray}

\section{Some mathematical manipulation}
\label{Ap_b}
In this appendix we present the procedure adopted to express the Casimir energy, given in \eqref{cas_energ}, in the more workable expression, Eq. \eqref{Energy_2}.  
Let's consider the integral in the following form:
\begin{eqnarray} \label{int_apend}
    \int_0^{\infty} kdk \int_{\sqrt{k^2+\mu^2}}^{\infty} f(u) \sqrt{(u^2-k^2)^{\xi}-\mu^{2\xi}} du .
\end{eqnarray}

The first step will be to perform a change of coordinates, $x=\sqrt{u^2-k^2-\mu^2}$. Thus, the integral above can be rewritten as:
\begin{eqnarray}
    \int_0^{\infty} kdk \int_0^{\infty} x \frac{f(\sqrt{x^2+k^2+\mu^2})}{\sqrt{x^2+k^2+\mu^2}} \sqrt{(x^2+\mu^2)^{\xi}-\mu^{2\xi}} dx .
\end{eqnarray}
Next, we should adopt a transformation to polar coordinates, $x=r\cos \theta$ and $k=r\sin \theta$, which gives us:
\begin{eqnarray}
    \int_0^{\infty} r^3 \frac{f(\sqrt{r^2+\mu^2})}{\sqrt{r^2+\mu^2}} dr \int_0^{\pi/2} \sin \theta \cos \theta \sqrt{(r^2\cos^2\theta + \mu^2)^{\xi}-\mu^{2\xi}} d\theta .
\end{eqnarray}

Finally, considering $w=\cos \theta$ and then $t=w^2$, we can represent the integral \eqref{int_apend} as:
\begin{eqnarray}
    \frac{\mu^{\xi}}{2} \int_0^{\infty} r^3 \frac{f(\sqrt{r^2+\mu^2})}{\sqrt{r^2+\mu^2}} dr \int_0^1 \sqrt{\left({r^2t}/{\mu^2} + 1\right)^{\xi}-1} dt   \  .
\end{eqnarray}

\end{document}